\documentclass[12pt]{article}
\usepackage{graphicx}
\usepackage{amsmath,amssymb}
\newcommand{\dd}{\mbox{d}}
\begin{document}
\title{Two-photon physics at future electron-positron colliders}
\author{I.R.Boyko, V.V.Bytev
and A.S.Zhemchugov\\
Joint Institute for Nuclear Research, Dubna, Russia
}

\maketitle

\abstract{
Two photon collisions offer a variety of
physics studies that can be performed
at the future electron-positron colliders.
Using the planned CEPC parameters as a benchmark
we consider several topics to be studied
in the two-photon collisions.
With the full integrated luminosity the Higgs boson
photoproduction can be reliably observed.
A large statistics of various quarkonium states
can be collected. The LEP results on the photon
structure function and the tau lepton anomalous
magnetic moment can be improved by 1-2 orders
of magnitude.
} 

\maketitle

\section {Introduction}

The flagship of the modern particle physics, the LHC,
is expected to continue running for more than 10 year
from now. However, the planning of the next generation
colliders has already started. 

It is widely believed that the next major collider project
can be a high-energy, high-luminosity electron-positron collider.
Currently four mature projects of $e^+e^-$ colliders 
are under consideration: linear colliders CLIC \cite{CLIC}
and ILC \cite{ILC} and circular colliders CEPC \cite{CEPC}
and FCC-ee \cite{FCCee}. 

The scientific programs of the future $e^+e^-$ colliders
are well developed.
The main goals are the Higgs boson physics; high-precision measurements 
at the Z pole energy; the top quark physics; searches
for new physics phenomena. In this paper
we investigate prospects for another branch 
of the experimental program: the physics 
of two photon collisions.

The collisions of (quasi-)virtual photons are responsible for a significant 
fraction of the total event rate at electron-positron
colliders. They are especially abundant at the
linear colliders due to high acceleration gradients.
Such events represent an unpleasant background 
for the studies of other physics processes.
On the other hand, the two-photon collisions
themselves provide opportunities for interesting
physics studies. In this paper we analyze
several topics which can be investigated
using the two-photon collisions at the
future $e^+e^-$ colliders.

At the linear colliders the rate and the energy spectrum 
of the two-photon collisions depend of the particular
configuration of beams. Therefore our study is restricted
to the case of circular colliders, where the 
differential gamma-gamma luminosity can be predicted
from the first principles.

Both CEPC and FCC are expected to take large data sets
at 240 GeV center-of-mass energy (a point close
to the maximum yield of Higgs bosons). The CEPC expected
integrated luminosity is 5.6 ab$^{-1}$. FCC plans
to collect at least 5 ab$^{-1}$ at 240 GeV, followed
by 1.7 ab$^{-1}$ at the energies of top pair production
(340-365 GeV). In this paper we use the CEPC planned
energy and luminosity as a benchmark scenario; 
however, our qualitative results are also applicable
to the FCC which has a similar expected performance.

\section {Two-photon collisions at electron-positron colliders}

The era of two-photon physics had been started in early 1970th, 
when a relatively high cross section of the 4-th order QED process 
$e^+e^-\to e^+e^-e^+e^-$ was observed  \cite{Balakin:1971ip}
in Budker Institute of Nuclear 
Physics (Novosibirsk).

It should be noted that the first theoretical predictions for 
a photoproduction of  pions 
\cite{Low:1960wv}, \cite{Calogero:1960zz} in $e^+e^-$ collision had been 
made back around 1960, the calculated cross-sections deemed unmeasurably 
small and no further elaboration has been done till the late 60th. 
At that time a series of papers covering multiple final states of 
$e^+e^-$ colliders had been published, but only after experimental 
observation at   Novosibirsk the   most remarkable mechanism 
of cross-section enhancement due to
small virtuality of photons emitted by the  electrons 
was revealed \cite{Ginzburg:2015saa}. 

It is the small virtuality of intermediate photons that features two main 
characteristics of two-photon processes: enhancement of the yield with the 
increase of  colliding beam energies and the possibility to study new type 
of  processes  $\gamma\gamma\to X$ with two variable parameters - 
the virtualities of intermediate photons.

The two-photon process can occur at electron-positron
or electron-electron colliders:
$$e^\pm+e^-\to e^\pm+e^- +\gamma^*+\gamma^*\to e^\pm+e^- +X,$$
where $X$ describes an arbitrary final state.
Not any final state can be produced in the two-photon processes:
since two photons have even C parity, only states with C=+1
are possible. As it was mentioned above, two-photon  cross-section 
slowly rises with the energy and becomes greater than the one of the 
annihilation channel at $\sqrt{s}/2 \gtrsim 1$ GeV.

\begin{figure}[htbp]
        \label{twophotonfig}
	\begin{center}
		\includegraphics[width=0.45\textwidth]{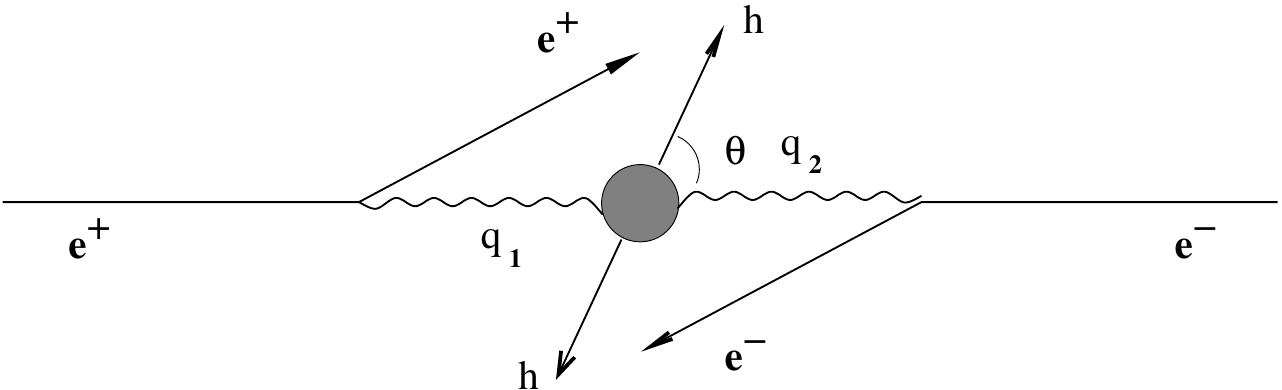}
	\end{center}
	\caption{Kinematics of a two-photon process.}
\end{figure}

In a two-photon process the initial particles (electrons and/or positrons) 
emit two photons with virtualities $q_1$ and $q_2$ and the latter  merge 
in some final system of particles X with invariant mass square 
$W^2={(q_1+q_2)^2}$, see Fig.\ref{twophotonfig}. 

The cross-section of two-photon scattering of electron (positron) can be 
calculated as a convolution of amplitudes describing the emission of 
virtual 
photons off initial particle and the $\gamma\gamma\to X$ transition. The 
first one is calculated within the QED, and the second can be 
expanded in independent tensors, the choice of which is quite arbitrary 
up to the conservation of the Lorenz invariance, T- and gauge invariance. 
For the sake of  physical interpretation, it is  encoded in five 
structure functions. Three of them can be expressed through the cross 
section $\sigma_{a,b}$ for  scalar ($a,b=S$) and transverse photons
($a,b=T$). The other structure functions $\tau_{TT}$
and $\tau_{TS}$ correspond to transitions with spin-flip for
each of the photons with total helicity conservation \cite{Budnev:1974de}:

\begin{eqnarray}
\dd\sigma=
\frac{\alpha^2}{16\pi^4q_1^2q_2^2}\sqrt{\frac{(q_1q_2)^2-q_1^2q_2^2}{(p_1p_2)^2-m_e^2m_e^2}}
\biggl(
4\rho^{++}_1\rho^{++}_2\sigma_{TT}
+2|\rho_1^{+-}\rho_2^{+-}|\tau_{TT}\cos(2\tilde{\phi})
\nonumber \\
+2\rho^{++}_1\rho^{00}_2\sigma_{TS}
+2\rho^{00}_1\rho^{++}_2\sigma_{ST}
+\rho^{00}_1\rho^{00}_2\sigma_{SS}
-8|\rho^{+0}_1\rho^{+0}_2|\tau_{TS}\cos(\tilde{\phi})
\biggr)
\frac{\dd^3 p_1'\dd^3p_2'}{E_1 E_2},
\label{budnevEq}
\end{eqnarray}

\noindent where $\rho_{i}^{ab}$ are the density matrices of the
virtual photon in the $\gamma\gamma$-helicity basis, $p_i, (p_i')$ corresponds to the momentum  and energy of 
initial (scattered) leptons, $ E_i$ stands for scattered lepton energies, $\tilde{\phi}$ is the angle between the 
scattering planes of colliding particles in c.m.s. of photons, $m_e$ is 
the mass of the initial lepton. 

The exact formula (\ref{budnevEq})  provides an accurate estimation of 
two-photon process at any kinematical region and could be used for 
quantitatively correct simulations \cite{Schuler:1997ex,Schuler:1997yw}. 
At the limit $q_i^2\to 0$ one can see a logarithmic enhancement of 
cross-section, with natural kinematic "regularization" value at $m_e^2$.

The small $q^2_i$ domain gives the main contribution to the cross-section 
of the process under consideration. In this limit the expression for 
(\ref{budnevEq}) can be simplified under the procedure called  Equivalent 
Photon Approximation (Weizacher-Williams' method).

Keeping in mind this approximation  and taking into account only leading terms
we can see that the cross-section of the two-photon process factorizes
into two different parts: one part is connected with emission of
two real photon by initial particles and the other is
the final state production by the two photons \cite{Brodsky:1970vk}:

\begin{eqnarray}
\frac{\dd\sigma^{(0)}}{\dd W^2\dd\Gamma}=2\biggl(\frac{\alpha}{\pi}\biggr)^2
\frac{1}{W^2}\ln^2\frac{E}{m_e}
f(\frac{W}{2E})\frac{\dd\sigma_{\gamma\gamma\to X}(W^2)}{\dd\Gamma},
\label{eqEPA}
\end{eqnarray}
where E is the energy of the initial particle and 
\begin{eqnarray}
\nonumber \\
f(\gamma)=-(2+\gamma^2)^2\ln\gamma-(1-\gamma^2)(3+\gamma^2)
\label{factorF}
\end{eqnarray} 
is the factor describing  the luminosity dependence over
the invariant mass of the colliding photons, $\Gamma$ stands for  phase volume of final state $X$.

The cost of this simplicity is 
the underestimation of cross-section in some
specific kinematics, the loss of the information on the initial particle 
scattering angle and missing deep virtual scattering behaviour of 
cross-section, when one of the photon has
a rather large invariant mass.
Of course, events with big virtuality of an intermediate photon
are much more rare, since there is no $q_i^2$ pole enhancement.

\begin{figure*}[h]
\vspace*{-25mm}
\begin{center}
\mbox{
\hspace*{-8mm}
\includegraphics[width=0.5\textwidth]{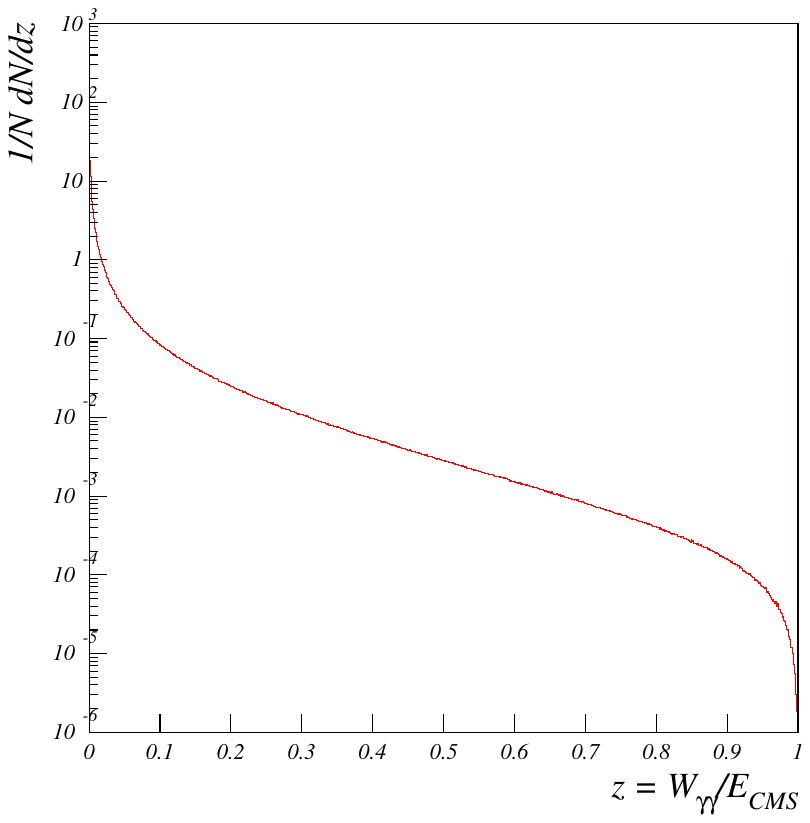}
\hspace*{-22mm}
\includegraphics[width=0.5\textwidth]{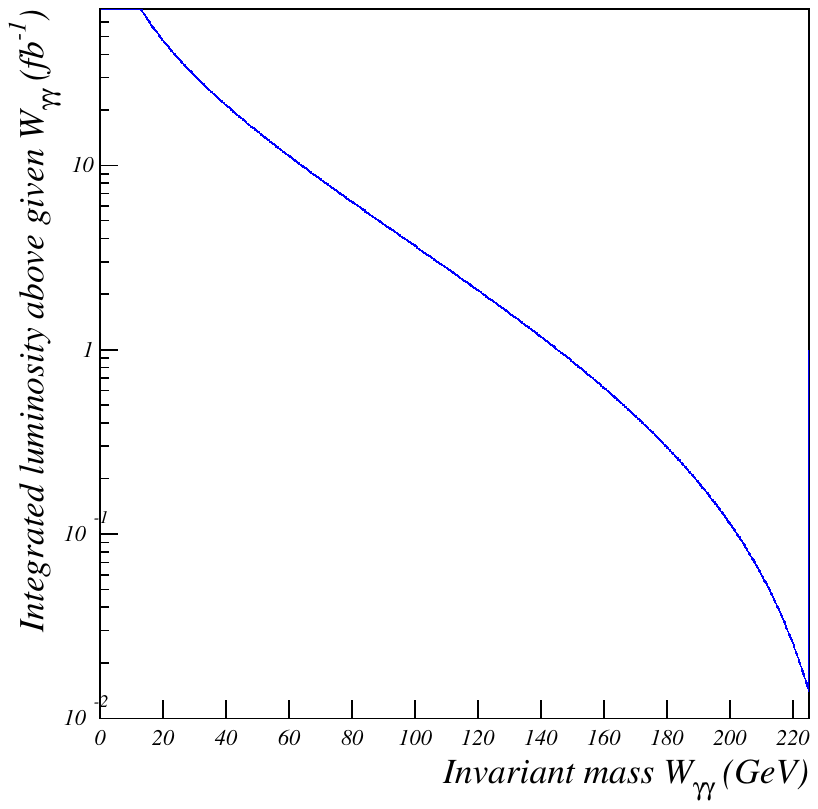}
}
\end{center}
\vspace*{-30mm}
\caption{
Left: number of $\gamma\gamma$ collisions
as a function of the $\gamma\gamma$ invariant mass relative 
to the $e^+e^-$ CMS energy. Right: integrated luminosity
of the $\gamma\gamma$ collisions above certain collision energy,
under the assumption of 5.6 ab$^{-1}$ integrated luminosity
of $e^+e^-$ collisions.
}
\label{gglum}
\end{figure*}

Small $q^2_i$ invariant mass of emitted photons means that the scattered 
leptons proceed undetected (so called no-tag events), but due to large 
cross-section of the process the no-tag events can be selected
even without full final state recovery \cite{Abrams:1979xk},
using the requirements of the small total transverse momentum
of the produced system and small value of its invariant mass ($W\ll 2E$).
The gamma-gamma collisions tend to occur at the c.m.s. energies
much less than the nominal collider energy, as can bee seen
in Fig. \ref{gglum} (left). Nevertheless, the amount of collisions at
rather high energies is also significant and can be used
to study different aspects of the two-photon physics.

Practically, one can tag two photon events by detection of one or both
scattered electrons (single
tag or double tag mode). Tagging allows  one to suppress background
significantly at the cost of steep reduction of available statistics.
In this case the requirement of tagging the lepton(s)  in a given
energy and angular range means non-vanishing virtualities of photon mass  
and gives one the possibility to measure the $q_i^2$ dependence of the
two-photon cross-section (\ref{budnevEq}). In this paper we 
suppose that scattered particles can be tagged in the luminosity 
monitor. A typical CEPC and FCC-ee acceptance down to 30 mrad 
is assumed.

The two-photon processes were always very interesting for physicists
due to the fascinating opportunity to study conversion of pure light to
matter. In this paper we cover rather conservative   and 
well-established topics 
of the two-photon physics, namely quarkonium  spectroscopy, Higgs 
production, tau pair production and photon hadronic structure.  All 
suggested  measurements  are aimed to improve significantly the precision 
of existing experiments and without any doubt are achievable under 
the planned 
characteristics of the future $e^+e^-$ colliders. 
The integrated luminosity   for the gamma-gamma 
collisions above the $W^+W^-$ threshold (see  Fig. \ref{gglum}, right) 
will be  slightly less than 1 fb$^{-1}$ which is comparable to
the total $e^+e^-$ luminosity recorded by the LEP2 experiments.

\section {Quarkonium  spectroscopy}
\label{qarkSect}
Quarkonium is regarded as the simple  hadronic systems to explore the QCD aspects at the
low energy regimes through its spectroscopy \cite{Eichten:1978tg}.

Quarkonia, the bound states of a heavy quark c, b and
the corresponding antiquark, can be most effectively studied 
at  $e^+e^-$ colliders. 
Since the discovery of the $J/\psi$  in 1974, 
a lot of information about bounded $b\bar b$ and $c \bar c$ states 
was elaborated.

Quarkonium can be described as bound quark-antiquark state, 
with Coulombic  short-distance potential   with logarithmic 
modification of coupling strength and linear long-distance potential 
for quark confinement description \cite{Eichten:1978tg},\cite{Eichten:2007qx}. 

Although there exist a lot of different ways to quantitatively  
describe quarkonia, e.g. 
Lattice QCD  \cite{Na:2012kp},  NRQCD 
methods \cite{Brambilla:1999xf}, Light front 
quark model\cite{Choi:2007se} or exotic like  
instanton liquid model \cite{Pandya:2019sma}, 
there are still many inconsistencies between predicted and measured 
radial excitation mass spectra of quarkonium states.

Due to negative charge parity of the photon, 
only  neutral particles with even charge conjugation $C=1$
can be produced in the two-photon collisions.   
There is a lot of interesting tasks concerning light mesons constituted 
from light $u, d$ quarks, e.g. $\pi$, $\eta$, $\eta'$ and their 
excited states, but we expect that CEPC experiments
will be insensitive to the hadronic systems 
with masses below 3 GeV due to the issues 
of detector resolution and experimental environment.
Therefore the most straightforward studies can be done 
with heavy $c$ and $b$ quarks, which lead us to the  
charmonia and bottomonia spectroscopy.

\begin{table*}[h]
	\begin{center}
		\begin{tabular}{ c| c| c|c |c|c|c}
			fraction& $ \eta_c $ &$ \chi_{c_ 0}$  & 	$ \chi_{c_ 1}$  &	$ \chi_{c_ 2}$  &  	$ \eta_c(2S)$ \\
			\hline
	$10^{-4}$	& $1.57\pm 0.12$ & $2.04\pm 0.09$ & $<6.3 \times 10^{-2}$ &$2.85\pm0.1$ &  $1.9\pm 1.3$ \\

		\end{tabular}
	\end{center}
	\caption{Charmonium  low state radiative decay branching ratio }
	\label{table:TwogamWidth}
\end{table*}

Charmonium states with even charge conjugation $C=1$ 
up to the $\eta_c(2S)$ mass of 3637 GeV are well studied, 
their radiative decays to photons have been seen and measured. 
 $ \chi_{c_ 1}$ production in two-photon collision  was recently seen by BELLE collaboration \cite{Belle:2020ndp}. However, only upper limit on branching ratio is available so far (see Table \ref{table:TwogamWidth}).
 From numerical 
estimation of event number (see Sect. \ref{GGsectEstim}, 
double tag set-ups) we can conclude that it is definitely possible 
to improve accuracy of  $\gamma\gamma$ branching ratio for 
$ \chi_{c_ 1}$ and 	$ \eta_c(2S)$ charmonium states, 
and to measure at least the $Q^2$ dependence of resonance 
formation in single-tag set-up, where we expect 
approximately $10^4$ events for each charmonium state. 

The event yield estimation of charmonium production with the mass larger 
than that of $ \eta_c(2S)$ is more tricky. From one point of view 
the $c\bar c$ spectroscopy at that region of mass states is very 
interesting due to lack of knowledge about internal structure 
and branching ratio to two photons of resonances 
(see Table \ref{table:TwogamPurpose}), so any results on 
measurement or lower limit estimation of $\gamma\gamma$ width 
are highly appreciated. Moreover, the measurement of $Q^2$ dependence 
of resonance formation could give us hard restriction over quantum 
numbers and internal structure of resonances under consideration.

The charmonium  bound states  with masses above 
open charm threshold can not be described only 
in the frames of  constituent quark model that  describes 
the observed meson spectrum as $q\bar q$ bound states 
(see Table \ref{table:TwogamPurpose}).  
To explain unexpected  quantum numbers, 
masses, branching ratios  and other properties of the heavy  
resonances that form the charmonium spectrum, various 
possibilities of new physics states are considered: meson states 
that are made of bound gluons (glueballs),  qq-pairs with an excited 
gluon (hybrids),  multiquark color singlet states such as qqqq 
(tetraquarks) and  molecular  bound states of 4-quark system,   
six-quark and 'baryonium' bound states.

The other side of the problem is that without a knowledge of the internal 
structure it is not possible to make a firm estimation of the charmonium 
state production. Even a naive equivalent photon approximation with  
narrow resonance approximation is not applicable due to unknown 
branching ratio to two photons.

A rough estimation of production rate we suggest to make under 
assumption that all charmonium states mostly consist of  $c\bar c$ states. 
We know that branching ratio of the two photon  quarkonium decay at lowest  
order does not depend on the state mass and has slight
(up to a factor of few) dependence on its quantum numbers  
(see \cite{Kwong:1987ak} and Table \ref{table:TwogamWidth}).  
So  two photon branching ratio could be estimated at the  level of $10^{-4}$. 
That estimation gives us about $10^7-10^8$ events in no-tag mode 
(although we do not observe scattered leptons, it is possible to reduce 
background events  by  imposing a strict transverse-momentum balance along 
the beam axis for the final-state hadronic system \cite{Uehara:2007vb}), 
and about $10^4$ events in single (minimal angle of detection at 6 degrees) 
and double tag (using the luminosity calorimeter) modes. 
All estimations are made with accelerator parameters described 
in Sect. \ref{GGsectEstim}. For the estimations with no-tag mode 
the events with scattered electrons were also accepted, 
with negligible contribution to the total event yield.

As a result we can conclude that if the recently discovered charmonium 
resonances (Table \ref{table:TwogamPurpose}) are similar to the charm 
quark bound state, we should definitely discover their  quantum numbers 
and measure the two photon branching ratios. 

\begin{table*}[h]
	\begin{center}
		\begin{tabular}{ c| c| c|c |c}
			name& $ J^{PC}$ & Width (MeV)   & $	\Gamma_{\gamma\gamma}$  &	nature   \\
			\hline
			$\chi_{c0}(3860)$ &$0^{++}$     &$ 201\pm101$& seen & $c\bar c$ +possible non-$q\bar q $ states \\
			$\chi_{c1}(3872)$ &$ 1^{++}$      & $1.19\pm0.21$     &seen& candidate for an exotic structure\\
			X(3915)           &0 or $2^{++}$& $20\pm 5$  &seen  &  $c\bar c$ +possible non-$q\bar q $ states \\
			$\chi_{c2}(3930)$ &$2^{++}$&$24\pm6$  &seen  &  $c\bar c$ +possible non-$q\bar q $ states \\
			X(3940)&$?^{??}$&$37\pm17$&-&  $c\bar c$ +possible non-$q\bar q $ states \\
			X(4050)&$?^{?+}$&$82\pm28$&-&   candidate for an exotic structure\\
			X(4100)&$?^{??}$&$152\pm70$&-&   candidate for an exotic structure\\
			$\chi_c1$(4140)&$1^{++}$&$22\pm7$&not seen&   candidate for an exotic structure\\
			X(4160)&$?^{??}$&$139\pm60$&-&   $c\bar c$ +possible non-$q\bar q $ states \\
			X(4250)&$?^{?+}$&$177\pm70$&-&   candidate for an exotic structure\\
			$\chi_{c1}$(4274)&$1^{++}$&$49\pm12$&-&   candidate for an exotic structure\\
			X(4350)&$?^{?+}$&$13\pm10$&seen&  $c\bar c$ +possible non-$q\bar q $ states \\
			$\chi_{c0}$(4500)&$0^{++}$&$92\pm29$&-&  candidate for an exotic structure\\
		    $\chi_{c0}$(4700)&$0^{++}$&$120\pm50$&-&  candidate for an exotic structure\\
		\end{tabular}
	\end{center}
	\caption{Charmonium  exotic states that could be seen at $\gamma\gamma$ collision  }
	\label{table:TwogamPurpose}
\end{table*}

For the bottomonium states only ground state $\eta_b(9399)$ decay to two 
photons was seen, and other $C=+1$ states radiative decay were not
measured \cite{Tanabashi:2018oca}, so any information about them,
e.g. upper limit on the two photon branching  or  decay event observation, 
are highly appreciated.

At the Sect. \ref{GGsectEstim} we estimate low bottomonium states $1S, 1P,2S$. 
Event number of  positive charge parity states with higher  radial quantum 
number, namely $\chi_{b0}(2P)$,  $\chi_{b1}(2P)$, $\chi_{b2}(2P)$, 
$\chi_{b1}(3P)$, $\chi_{b2}(3P)$    are  estimated as counterparts with   
radial quantum number equal to unity.

With the number of  bottomonium state events of approximately $10^5$ 
in no-tag mode and $10^2$ with double-tag we could conclude 
that we definitely can at least put upper limit on radiative 
decay branching of bottomonium states.

\subsection {Theoretical estimation}

We can estimate theoretically the full cross-section of two photon 
meson production with the help of 
equivalent photon approximation formula (\ref{eqEPA}),
where cross-section of meson with spin J, mass M  photoproduction 
can be estimated trough:
\begin{gather}
\sigma_{mes}(W^2)=(2J+1)8\pi^2 \frac{\Gamma_{\gamma\gamma}}{M}
\frac{1}{\pi}\frac{M\Gamma}{(M^2-W^2)^2+M^2\Gamma^2},
\end{gather}
here $\Gamma_{\gamma\gamma}$ and  $\Gamma$ are the two photon and total width of meson decay.

\subsection {Number of event estimation}
\label{GGsectEstim}

Simulation was performed with the GALUGA 
\cite{Schuler:1997ex,Schuler:1997yw,Schuler:1996qr},
two photon production  generator, 
where virtualities $Q^2$ of photons are fully taken into account through 
five structure functions describing  photon scattering with different 
polarizations. Such a consideration gives one the possibility  to make 
a robust estimation  of meson production at large photon virtualities 
(large electron  or positron scattering angle).   
Resonance formation in two-photon scattering are calculated in the framework 
of constituent-quark model \cite{Schuler:1997yw}. 

The expected event yields are presented  for  the total beam 
energy $\sqrt{s}=240$ GeV and integrated luminosity $5$~ab$^{-1}$.

\begin{table*}[h]
\begin{center}
	\begin{tabular}{ c| c| c|c |c|c|c}
		name& No Tag & S Tag 10  & S Tag 6  & D Tag 6   & D Tag 6-1.9& D Tag 1.9-1.9  \\
		\hline
		$ \eta_c $  &  $1.\times 10^9 $  &  $5 .3\times 10^5 $  &  $1 .7\times 10^5$   &  $7 .6\times 10^2$   &   $1 .2\times 10^4$   &  $8 .6\times 10^4$ \\
		$ \chi_{c_ 0}$  &  $2.\times 10^8 $  &  $6 .5\times 10^4 $  &  $2 .2\times 10^4$   &  $6 .3\times 10^1$   &    $9.\times 10^2$   &  $8 .3\times 10^3$ \\
		$ \chi_{c_ 1}$  &  $7.\times 10^6 $  &  $4 .4\times 10^4 $  &  $1 .1\times 10^4$   &  $1 .7\times 10^2$   &    $2 .4\times 10^3$   &  $1 .7\times 10^4$ \\
		$ \chi_{c_ 2}$  &  $9 .5\times 10^7 $  &  $3 .1\times 10^4 $  &  $9 .8\times 10^3$   &  $5 .4\times 10^1$   &    $7 .2\times 10^2$   &  $6 .7\times 10^3$ \\
		$ \eta_c(2S)$   &  $2 .4\times 10^8 $  &  $2 .1\times 10^5 $  &  $6 .4\times 10^4$   &  $4 .1\times 10^2$   &    $6.\times 10^3$   &  $4 .1\times 10^4$ \\

	\end{tabular}
\end{center}
\caption{Estimation of event number for charmonium states.
Singe tag, no tag and double tag set-ups are considered. Cuts on electron or positron scattering angle are 1.9, 6 and 10 degrees.}
\label{table:1}
\end{table*}

Quarkonium production was estimated in three different modes, 
namely no tag, where no final  scattered electron or positron detected (there is no  limitation on the polar angle of the scattered electron/position), 
single tag, where electron or positron are detected and double tag 
(both electron and positron are detected). The  minimal angles are taken  
equal to 6 or 10 degrees (various detector characteristics are considered), 
and 1.9 degrees for the case when scattered electron or positron can be 
detected in the luminosity calorimeter.

\begin{table*}[h]
\begin{center}
	\begin{tabular}{ c| c| c|c |c|c|c}
		name& No Tag & S Tag 10  & S Tag 6  & D Tag 6   & D Tag 6-1.9& D Tag 1.9-1.9  \\
		\hline
		$ \eta_b (9399) $  &  $1.3\times 10^6 $  &  $1.4\times 10^4 $  &  $3.9\times 10^3$   &  $1.1\times 10^2$   &  $1.\times 10^3$   &  $3.6\times 10^3$ \\
		$ \chi_{b_0}(1P)$  &  $9.6\times 10^4 $  &  $5.2\times 10^2 $  &  $1.4\times 10^2$   &  $2.7$   &  $3.\times 10^1$   &  $1.5\times 10^2$ \\
		$ \chi_{b_1}(1P)$  &  $3.9\times 10^3 $  &  $5.1\times 10^2 $  &  $1.3\times 10^2$   &  $6.6$   &  $5.3\times 10^1$   &  $1.4\times 10^2$ \\
		$ \chi_{b_2}(1P)$  &  $8.3\times 10^4 $  &  $4.9\times 10^2 $  &  $1.2\times 10^2$   &  $3.4$   &  $3.4\times 10^1$   &  $1.6\times 10^2$ \\
		$ \eta_b(9999)$   &  $4.6\times 10^5 $  &  $5.8\times 10^3 $  &  $1.6\times 10^3$   &  $4.9\times 10^1$   &  $4.4\times 10^2$   &  $1.5\times 10^3$ \\

	\end{tabular}
\end{center}

\caption{Estimation of event number for bottomonium states.
Singe tag, no tag and double tag set-ups are considered. Cuts on electron or positron scattering angle are 1.9, 6 and 10 degrees.}
\label{table:2}
\end{table*}

From the Tables (\ref{table:1},\ref{table:2}) one can see that the numbers 
of registered mesons are drastically dependent on the minimal angle
of the scattered
lepton detection. By using the luminosity calorimeter for detection 
of scattered leptons (the minimal detection angle of 1.9 degrees)  one can 
enhance the statistics of registered mesons by  two  orders of magnitude. 
This emphasizes the significance  low-angle calorimeter 
for the two-photon physics. 

As one can see, in no-tag mode, where no lepton are registered at final state, 
we expect a large number of events,  $10^7-10^8$ for quarkonium states and   
$10^5$ for bottomonium. But the absence of kinematical constraints
in the no-tag mode
will result in huge background contamination. 
Some special techniques like strict transverse-momentum balance 
for final state could be applied to reduce background.  
The exact estimation of no-tag quarkonium sensitivities could be  
produced only with a detailed detector simulation.

For quarkonium  two-photon physics on CEPC we suggest to use a conservative 
estimation, namely the single-tag mode (one detected lepton helps 
to reconstruct final state kinematics and drastically reduces the 
background events) and efficiency of detection about  $10\%$. 
It that case we will reconstruct about $10^2-10^4$ events for each 
charmonium state. For bottomonium states we could definitely  
measure low-state $\eta_b$ with number of registered events 
about $4\times10^2$ and possibly see some events or make upper bound  
for higher bottomonium states.

\section {Higgs boson production}
\label{higgsSect}

The vertex $H\gamma\gamma$ is forbidden in the SM
at tree level. The decay process $H \rightarrow \gamma\gamma$
as well as the production process $\gamma\gamma \rightarrow H$
proceed mostly through the top-quark  and W loops  and is  sensitive to
 contributions of new charged particles, so
an observation of an excess in the $\gamma\gamma \rightarrow H$
process would indicate a new physics phenomena,
e.g. a contribution of the anomalous $H\gamma\gamma$ vertex.

At $e^+e^-$ colliders the main background to the process
$\gamma\gamma \rightarrow H$ (referred as ``signal''
hereafter) are the Higgs boson
production via the fusion of virtual Z bosons
($ZZ\rightarrow H$) and $\gamma\gamma$ collisions
with final states identical to those of Higgs decays
but without a formation of the intermediate Higgs boson.
The background from $\gamma\gamma$ collisions
can be strongly suppressed by selecting the ``single-tag''
events where one of the beam particles is scattered
to a significant angle and is detected in the luminosity calorimeter.
The signal reduction due to this selection is relatively
small (by only a factor of 3 to 5) since the Higgs boson production
is characterized by a large $q^2$ transfer. In the following
we assume an event selection with a beam particle scattered
by at least 30 milliradians, which is well within the acceptance
of the luminosity monitor \cite{CDR}. The signal and the main
background sources have been simulated with PYTHIA generator
\cite{PYTHIA}. No detailed detector simulation has bee performed,
however the main features of the proposed CEPC detector
\cite{CDR} have been taken into account.

\subsection {Higgs photoproduction measurement plans}

Today the Higgs photoproduction at future lepton colliders 
attracts surprisingly low attention. 

Nevertheless there was a vivid discussion in frames of Higgs boson production
in ultraperipheral collisions (UPC) at LHC (proton and heavy ion  
collision cases was considered, see Silveira talk \cite{Ducati:2011ma}  
and references therein ) where  estimations of Higgs photoproduction was 
made for different set-ups and energies.

Recently there was published a paper \cite{Li:2019bnk} devoted to the estimation  of  photoproduction of Higgs boson at the LHeC \cite{AbelleiraFernandez:2012cc}, proposed electron-proton collider at LHC.

Also there exist a lot of estimations and proposals to measure double 
Higgs photoproduction at a photon collider,
particularly due to the possibility to probe trilinear Higgs interactions
\cite{Takahashi:2009sf},\cite{Belusevic:2004pz}.

\subsection {Theoretical estimation of event number}

For estimation of Higgs production rate due to two-photon mechanism we could utilize the equivalent photon approximation, 
elaborated in the papers  \cite{Low:1960wv},
\cite{Brodsky:1971ud}, and \cite{Budnev:1974de}:

$$\sigma_{ee\to eeH} 
\approx 2\biggl(\frac{\alpha}{\pi}\biggr)^2\biggl(\ln\frac{E}{m_e}\biggr)^2
\times
\int_0^{4E^2} \frac{\dd W^2}{W^2}f(\frac{W}{2E}) \sigma_{\gamma\gamma\to H}(W),
$$
\noindent here $\sigma_{\gamma\gamma\to H}(W)$ means Higgs photoproduction 
cross-section estimation and  function $f$ was defined in  eq. (\ref{factorF}).
 Here we consider only leading logarithm approximation.

The subprocess cross section for the two photon Higgs production can be 
calculated  trough narrow resonance estimation \cite{Brown}: 

$$
\sigma_{\gamma\gamma\to H}(W)\approx8\pi^2 \frac{\Gamma_{\gamma\gamma}}{M_H}\frac{1}{\pi}\frac{M_H\Gamma}{(M_H^2-W^2)^2+M_H^2\Gamma^2} \approx 
$$
$$
\approx 8\pi^2\frac{\Gamma_{\gamma\gamma}}{M_H}\delta(W^2-M_H^2),
$$
\noindent here $\Gamma_{\gamma\gamma}$ and $\Gamma$ are the two photon  and total width of Higgs decay correspondingly, $M_H$ is the Higgs mass.

Given the energy of initial  electron beam $E= 120$ GeV and 
$\Gamma_{\gamma\gamma}=2.27 \times10^{-3}\Gamma$, $\Gamma\approx 4.2 $ MeV  
\cite{Tanabashi:2018oca} we could roughly estimate the Higgs two
photon production at CEPC energies at the level of $0.25$ fb. 

It is very interesting that the Higgs production rate through 
two photon mechanism at CEPC and LHC are comparable. 
In the latter case the estimation gives the cross section about $0.1$ fb,
see \cite{Khoze:2001xm},\cite{Levin:2008gi},\cite{dEnterria:2009cwl}

\subsection {Background from the $ZZ$ fusion}

\begin{figure*}[h]
\vspace*{-25mm}
\begin{center}
\mbox{
\hspace*{-8mm}
\includegraphics[width=0.5\textwidth]{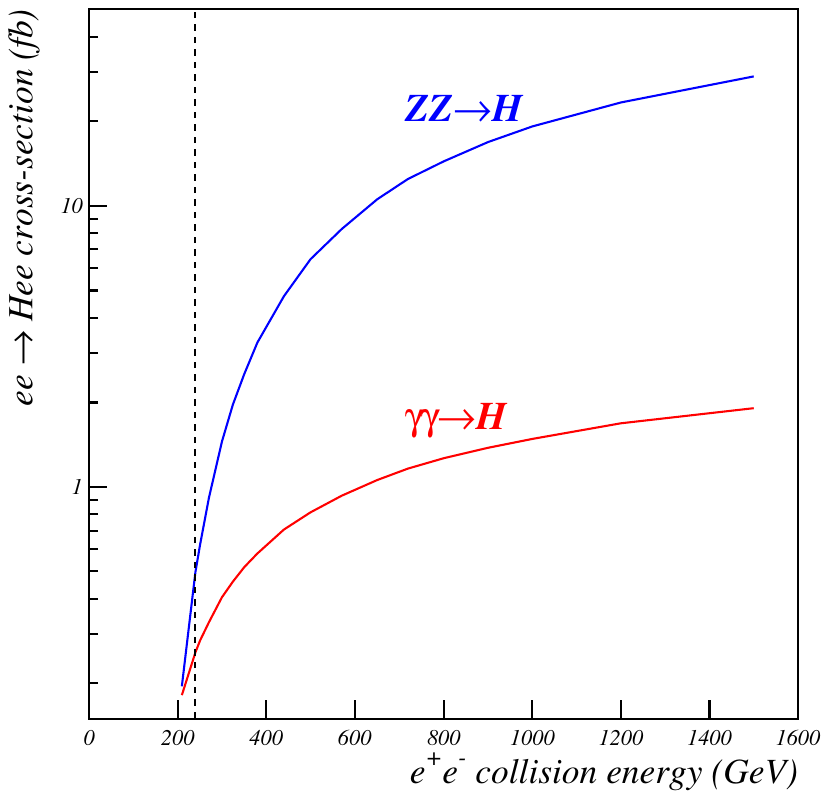}
\hspace*{-22mm}
\includegraphics[width=0.5\textwidth]{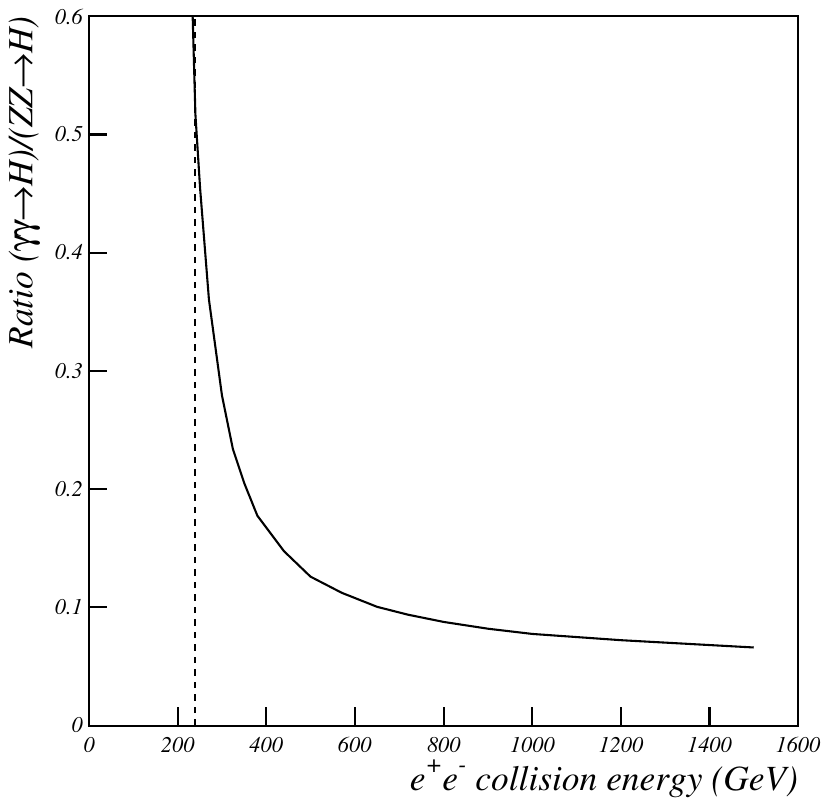}
}
\vspace*{-25mm}
\caption{
Left: energy dependence of $e^+e^- \rightarrow He^+e^-$
cross-section for the signal $\gamma\gamma \rightarrow H$
and background $ZZ \rightarrow H$ contributions.
Right: energy dependence of the signal to background ratio.
Vertical lines indicate 240 GeV CMS energy.
}
\label{figxsec}
\end{center}
\end{figure*}

The energy dependencies of the signal $\gamma\gamma \rightarrow H$
and the background $ZZ\rightarrow H$ processes \cite{PYTHIA}
are compared in Fig.\ref{figxsec}. One can see that
at high energy colliders the $ZZ\rightarrow H$ background
rate is much higher than that of the signal. However, at the $e^+e^-$
CMS energies near 240 GeV the background drops abruptly
and becomes comparable to the signal. The CEPC and FCC-ee colliders
are therefore perfectly suited to study
the $\gamma\gamma \rightarrow H$ process.

\begin{figure*}[h]
\vspace*{-25mm}
\begin{center}
\hspace*{-12mm}
\includegraphics[width=0.6\textwidth]{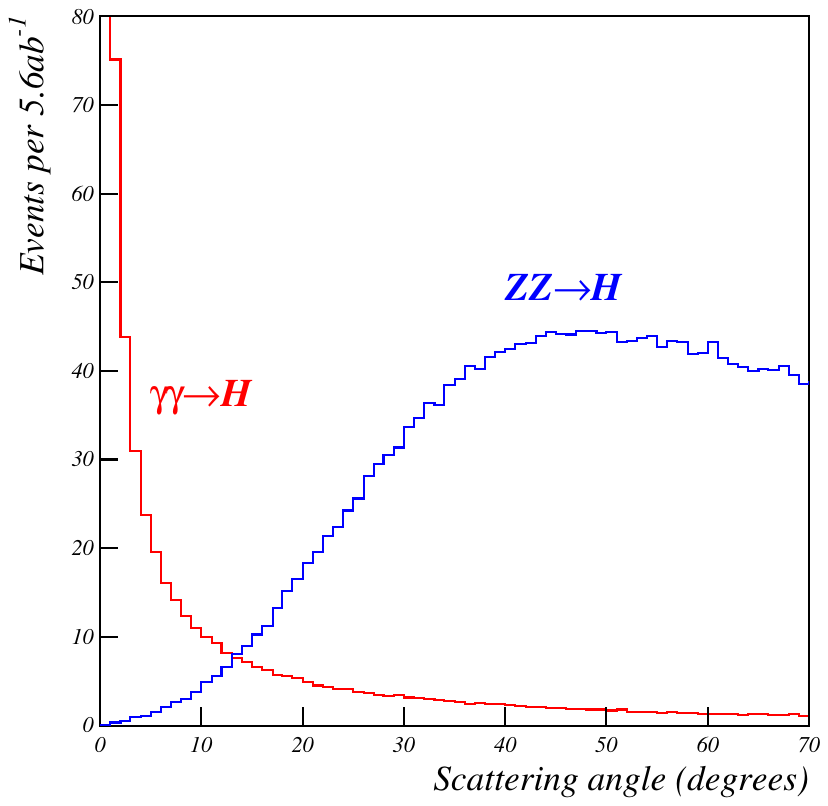}
\vspace*{-30mm}
\caption{
Distribution of the beam particle scattering angle
for $\gamma\gamma \rightarrow H$ signal and 
$ZZ \rightarrow H$ background.
The number of events is normalized to 5.6 ab$^{-1}$ integrated
luminosity.
}
\label{figth}
\end{center}
\end{figure*}

At 240 GeV the total signal cross-section is 0.26 fb \cite{PYTHIA},
to be compared with 0.50 fb for $ZZ \rightarrow H$ background.
The background can be significantly reduced
using the fact that the typical $q^2$ transfer
in the $ZZ \rightarrow H$ events
is much larger than
in the $\gamma\gamma \rightarrow H$ process.
The distribution
of the scattering angle is presented in Fig.\ref{figth}.
Nearly all background is removed with the requirement
that the beam particle is scattered by less than 24$^\circ$.

\begin{figure*}[h]
\vspace*{-25mm}
\begin{center}
\hspace*{-12mm}
\includegraphics[width=0.6\textwidth]{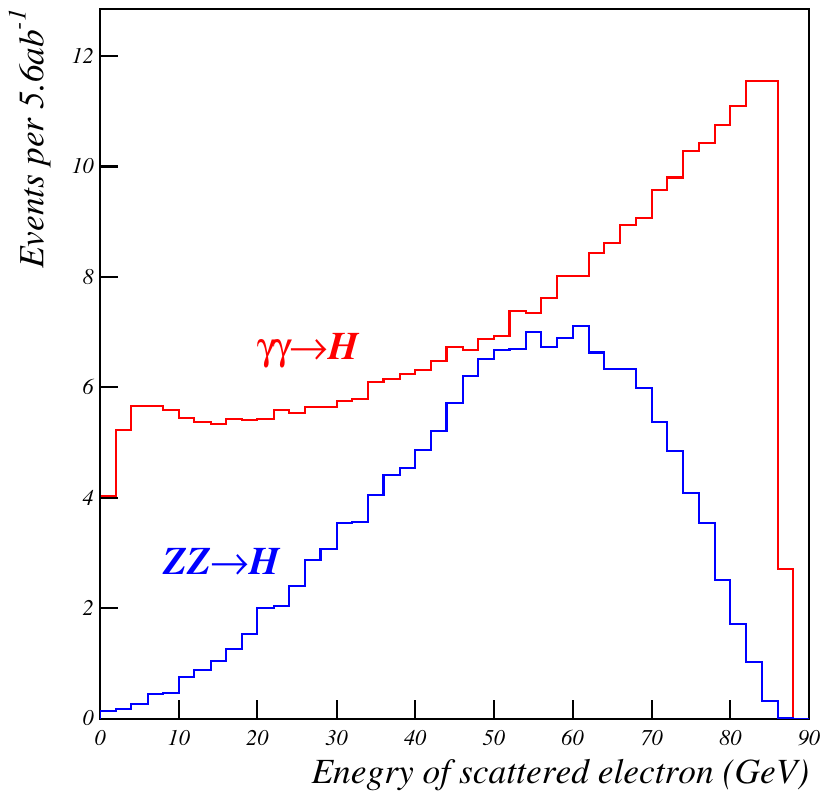}
\vspace*{-30mm}
\caption{
Energy distribution of the beam particle
with scattering angle between 30 mrad and 24$^\circ$
for $\gamma\gamma \rightarrow H$ signal 
and $ZZ \rightarrow H$  background.
}
\label{figpscat}
\end{center}
\end{figure*}

Fig.\ref{figpscat} shows the energy distribution
for the scattered beam particles with the scattering angle
between 30 mrad and 24$^\circ$. An additional energy cut
$E > 15$ GeV is applied to the scattered particle
to ensure a reliable identification in the
luminosity calorimeter. After the cuts
on the angle and the energy of the scattered particle
the cross-section is 0.049 fb for the signal
and 0.027 fb for the $ZZ \rightarrow H$ background.
Assuming 5.6 ab$^{-1}$ integrated luminosity, this corresponds
to 273 and 154 events, respectively.

\subsection {Background from other hard processes}

The Higgsstrahlung $e^+e^- \rightarrow ZH$ is the most
abundant Higgs production process at 240 GeV collision energy.
The ``single-tag'' signal can be caused
by an electron from the Z-boson decay.
The cross-section of $e^+e^- \rightarrow ZH \rightarrow eeH$
is 6.7 fb. A requirement that one of the electrons
is found in the ``single-tag'' region between 30 mrad and 24$^\circ$
reduces the effective cross-section to 1.4 fb.
Further background reduction is achieved by the requirement
that the second electron is not reconstructed
in the tracker. The events with second
electron outside of the tracking system
(less than 10 degrees from the beam)
correspond to 0.030 fb, which is similar to
the background from ZZ fusion. Still this electron
can be reconstructed in the luminosity calorimeter.
Although the reconstruction efficiency is not perfect, the background
can be reduced to a negligible level compared to ZZ fusion.
The electrons outside of the luminosity calorimeter
(less that 30 mrad from the beam) correspond
to 0.001 fb, which is also negligible.

Another ``standard'' mechanism of Higgs boson
production is the  WW fusion $e^+e^- \rightarrow H\nu\nu$.
The ``single-tag'' signal can be produced by an ISR photon
in the angular acceptance between 30 mrad and 10$^\circ$
(at larger angles photons are distinguished from electrons
by the tracking system). The total WW fusion cross-section
is approximately 5 fb at 240 GeV. The presence
of the ``tag'' photon reduces this to 0.032 fb,
which is similar to the ZZ fusion background.
Further background reduction is based on
the large missing transverse momentum in $H\nu\nu$ events.
A very loose cut $P_T^{miss} < 20$ GeV/c rejects
only 4\% of the signal while the WW fusion background
is reduced to 0.008 fb, which is only a small fraction
of the ZZ fusion background.

A W pair production is a potentially dangerous
background due to its high cross-section
(approximately 15 pb at 240 GeV). The ``single-tag''
signal can be produced by electrons from the leptonic W decays.
However, this background is reduced by several
very large factors: branching of $WW \rightarrow cse\nu$ decays
(factor 14); probability of simultaneous fake
b-tagging of both $c$ and $s$ quarks (factor of at least 500);
electron production outside of the tracking system
(factor 70); reconstruction of the hadronic W decay
with a Higgs boson mass (factor of at least 20);
requirement of a low $P_T^{miss}$ (factor of 7).
Taken together, the above factors reduce
the $e^+e^- \rightarrow W^+W^-$ to a negligible level.

\subsection {Background from non-resonant $\gamma\gamma$ collisions}

The most abundant (58\%) Higgs decay channel is $H \rightarrow bb$.
In this channel the most significant background is the non-resonant
production of b quark pairs in the $\gamma\gamma$ collisions,
$\gamma\gamma \rightarrow bb$. In addition,
there is a reducible background $\gamma\gamma \rightarrow cc$,
where both jets from c quarks are tagged as b jets.
The background from light quarks can be neglected
since it is efficiently suppressed by the b-tagging.

With the ``single-tag'' selection described above
the $\gamma\gamma \rightarrow bb$ cross-section
is 94 fb for the full range of the bb invariant masses.
It is reduced to 0.77 fb for $M_{bb} > $ 100 GeV.
According to the CEPC CDR \cite{CDR}, the 
invariant mass resolution in $H \rightarrow bb$ decays
is about 5 GeV. Within the $\pm 1\sigma$ window around the Higgs mass
($120 < M_{bb} < 130$ GeV) the $\gamma\gamma \rightarrow bb$
background is 0.124 fb, more than 5 times 
higher than the signal (taking into account the branching fraction
of the $H \rightarrow bb$ decay).

The total cross-section of the single-tag
$\gamma\gamma \rightarrow cc$ production
is 2086 fb. It is reduced to 13.6 fb
for $M_{cc} > $ 100 GeV and to 1.8 fb
for $120 < M_{cc} < 130$ GeV. Although the cc background
is almost 2 orders of magnitude higher than
the signal, it can be efficiently reduced by the b-tagging.
According the the CEPC CDR, the c jet rejection
factor is 10 for the 80\% b jet efficiency.
Applying the b-tagging to both jets, one gets
64\% efficiency for the signal tagging
and a factor of 100 reduction of the charm background,
making the latter much smaller than the 
$\gamma\gamma \rightarrow bb$ background.

\begin{figure*}[h]
\vspace*{-25mm}
\begin{center}
\hspace*{-12mm}
\includegraphics[width=0.6\textwidth]{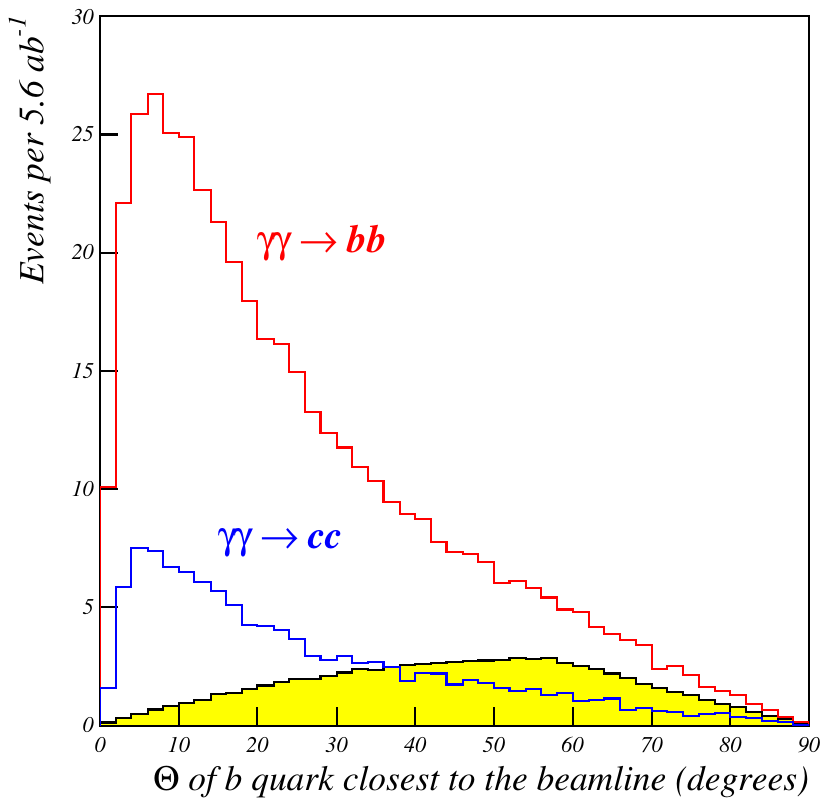}
\vspace*{-25mm}
\caption{
Distribution of the polar angle of the quark 
closest to the beam line. Shaded area is
the $\gamma\gamma \rightarrow H$ signal,
open histograms represent the non-resonant background.
The events are shown within the invariant mass 
window $118 < M_{bb} < 132$ GeV.
}
\label{figthb}
\end{center}  
\end{figure*}

The non-resonant background can be additionally reduced
using a cut on the direction of the produced b and c jets.
Fig. \ref{figthb} shows the distribution of 
the quark polar angle. The background jets are concentrated
near the beam axis. We apply a polar angle cut $\Theta > 20^\circ$ 
for both quarks. Within this acceptance we assume (for 
the signal and the background)
a 75\% efficiency to reconstruct both jets from b or c quarks.

\begin{figure*}[h]
\vspace*{-25mm}
\begin{center}
\hspace*{-20mm}
\includegraphics[width=0.6\textwidth]{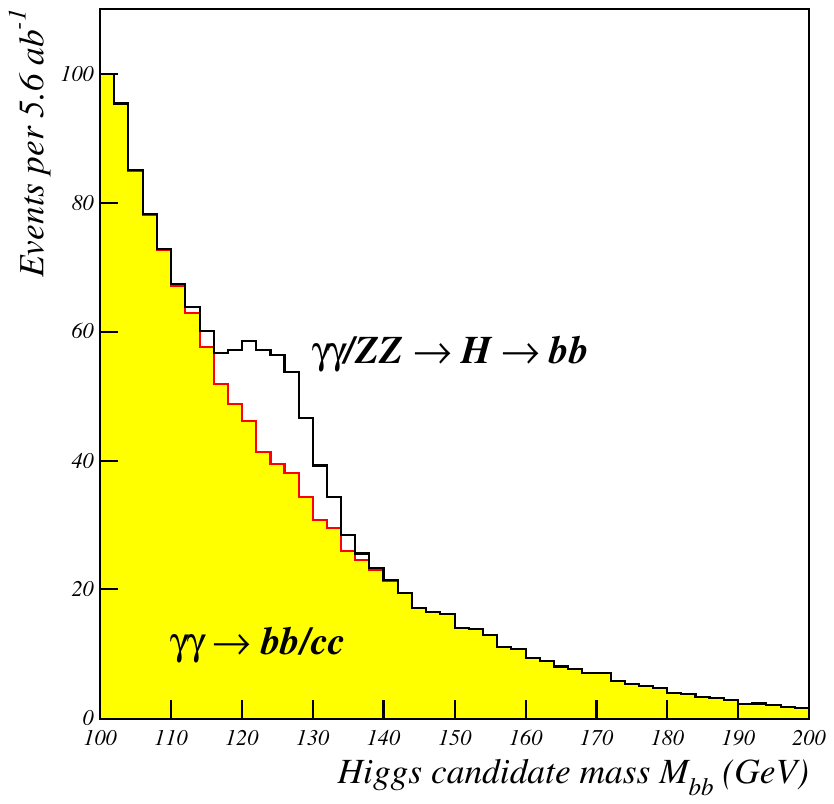}
\vspace*{-25mm}
\caption{
Distribution of the invariant masses of the Higgs candidates.
Shaded area represents $\gamma\gamma \rightarrow bb/cc$
backgrounds,
open histogram is the signal, with a small contribution
from the $ZZ \rightarrow H$ events.
}
\label{figmbb}
\end{center}
\end{figure*}

The $M_{bb}$ invariant mass distribution
is presented in Fig.\ref{figmbb}. 
The Higgs signal is smeared assuming 
a 5 GeV mass resolution. Within the window 118-132 GeV
the expected signal is 57 events, the peaking ZZ 
background is 33 events and the non-peaking 
background is 278 events.
A fit to the signal and background
yields results in about $4.1\sigma$ signal significance
for the $H \rightarrow bb$ channel, after the subtraction
of the $ZZ \rightarrow H$ background.
The signal significance can be further improved by including
other Higgs decay modes. We conclude that the  
$\gamma\gamma \rightarrow H$ signal can be reliably 
observed with the planned CEPC luminosity.

\section {Tau pair production}
\label{tauSect}

At present, the anomalous magnetic moments of electron
and muon are measured with an enormous precision,
better than one per billion and one per million, respectively.
These measurements provide an extremely important
test of the Standard Model. At the same time,
the anomalous magnetic moment of the tau lepton,
$a_\tau$, is known with rather poor accuracy.

The measurement of $a_\tau$ is interesting in two respects.
First, because of the large tau mass, $a_\tau$ is sensitive
to the contributions of the new physics at higher scales.
Second, many theoretical models predict that the new physics
effect manifest themselves only in the properties
of the third-generation fermions. Thus, the overwhelming
success of the Standard Model observed in the
sector of anomalous magnetic moment might be just
a consequence of performing the high precision measurements
with the leptons of only first generations.

The most precise determination of $a_\tau$ (17 permille)
was performed by the DELPHI experiment at LEP2.
The total luminosity was approximately 0.5 fb$^{-1}$
taken at the c.m.s. energies between 182 and 208 GeV.
The tau anomalous magnetic moment was extracted
from the absolute cross-section of the tau pair production
in gamma-gamma collisions. The simplest final state
was selected with one tau decaying to an electron,
another to a muon. The DELPHI precision of the cross-section
measurement was about 4\%.

At CEPC the integrated luminosity will be increased
by 4 orders of magnitude with respect to LEP2.
Thus, an improvement of $a_\tau$ precision by a large factor
can be expected.

At the collision energy of 240 GeV the QED cross-section
of the $e^+e^- \rightarrow e^+e^-\tau^+\tau^-$ process
is 570 pb \cite{BDKRC}.
This corresponds
to nearly 3 billions events with 5 ab$^{-1}$ of  integrated luminosity,
or 165 millions events with the $e-\mu$ final state.

At LEP2 the selection efficiency for the $e-\mu$ final state
was 15-20\%. In the CEPC environment a tighter selection
might be necessary to cope with the high background.
We conservatively
consider the following severe cuts: at least one of the two tracks
must have the transverse momentum $p_T > 5$ GeV/c,
the other track must have $p_T > 3$ GeV/c.
In addition, we require that directions of the both tracks
are more that 20 degrees from the beam axis,
and that the total energy of the two particles
is less than 30 GeV to remove the annihilation events.

\begin{figure*}[h]
\vspace*{-25mm}
\begin{center}
\mbox{
\hspace*{-8mm}
\includegraphics[width=0.5\textwidth]{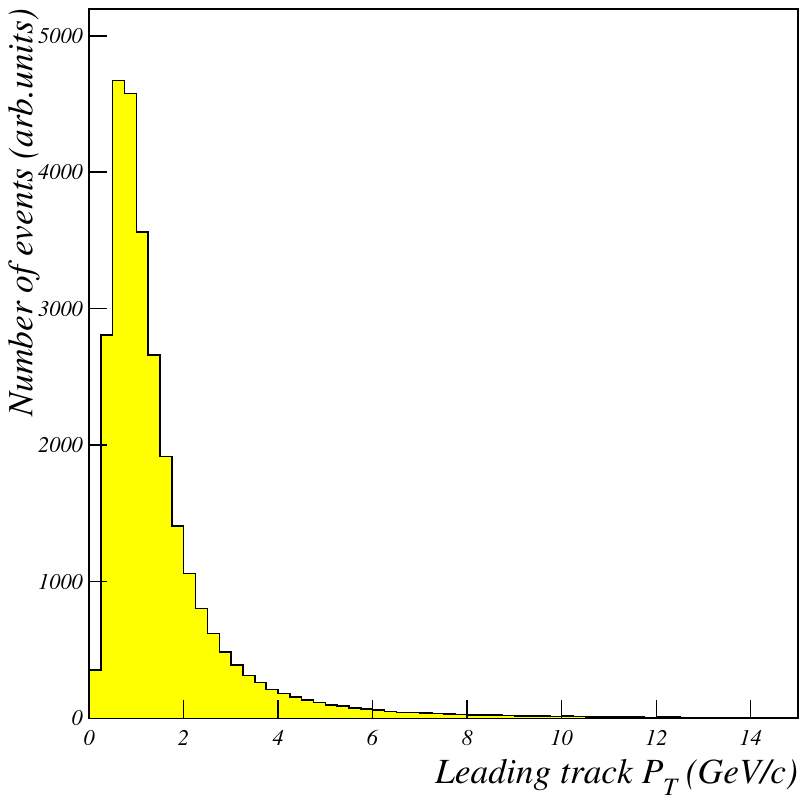}
\hspace*{-22mm}
\includegraphics[width=0.5\textwidth]{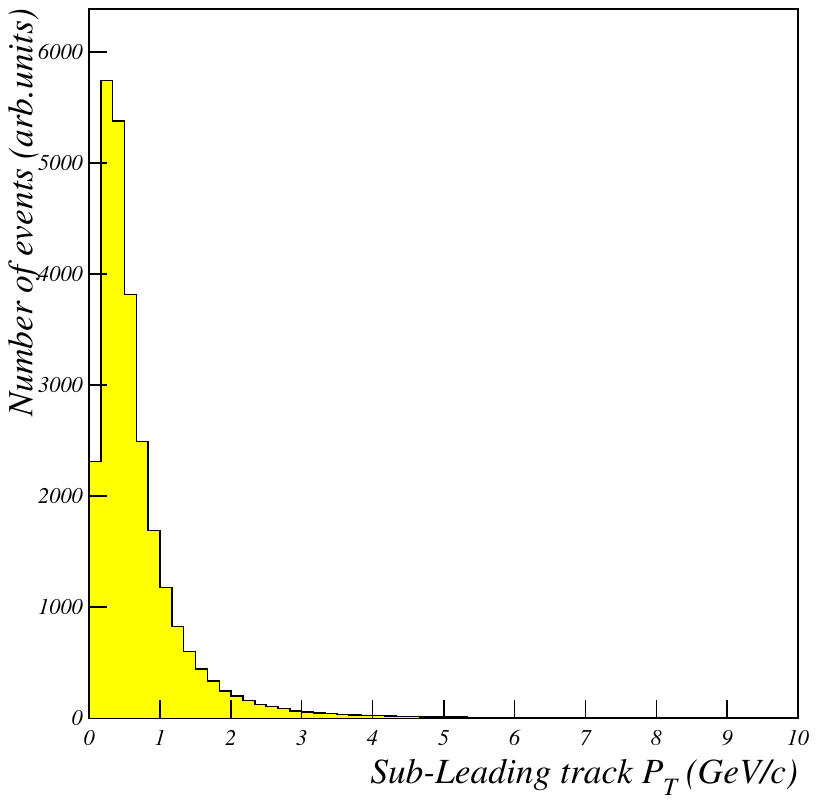}
}
\vspace*{-25mm}
\caption{
Distribution of the transverse momenta of the leading (left)
and subleading (right) tracks
in the $\gamma\gamma \rightarrow \tau\tau \rightarrow e\mu$ events.
}
\label{figptemu}
\end{center}
\end{figure*}

\begin{figure*}[h]
\vspace*{-25mm}
\begin{center}
\hspace*{-20mm}
\includegraphics[width=0.6\textwidth]{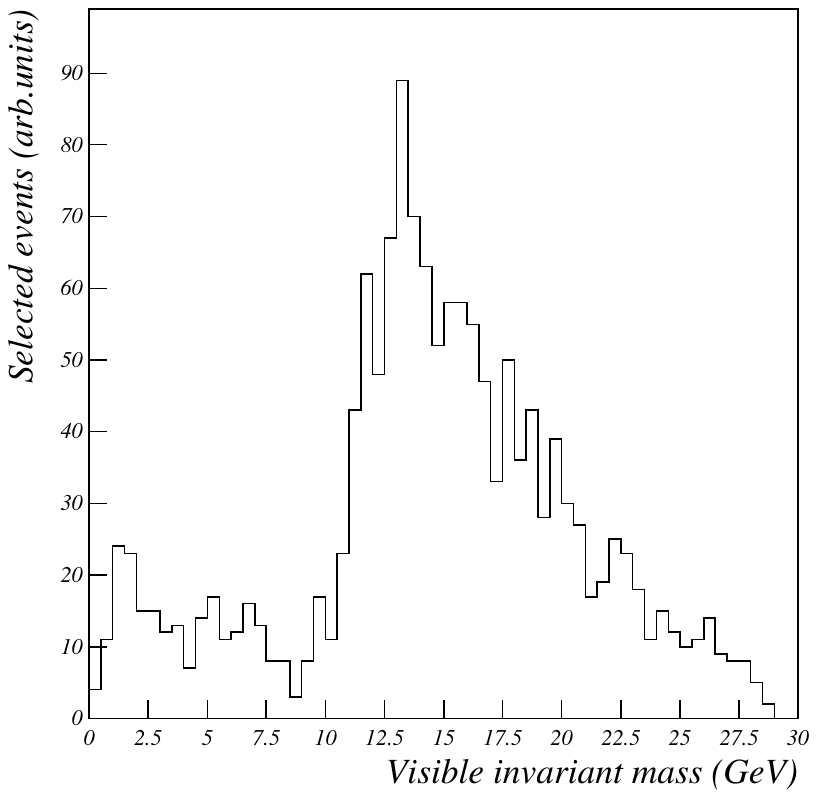}
\vspace*{-25mm}
\caption{
Distribution of the visible invariant masses of the
selected $\gamma\gamma \rightarrow \tau\tau \rightarrow e\mu$ events.
}
\label{figmvis}
\end{center}
\end{figure*}

Fig.\ref{figptemu} shows the distribution of the transverse
momentum of the leading and subleading tracks
in the $\gamma\gamma \rightarrow \tau\tau \rightarrow e\mu$ events.
One can see that only a very small fraction of the events
satisfy the selection cuts. Fig.\ref{figmvis} shows
the distribution of the invariant mass of the electron
and the muon after applying the selection cuts.

The above selection has the generator-level efficiency of 0.42\%,
corresponding to the total statistics of about
700 thousands events. Thus, the statistical error
is expected to be at the permille level and the measurement
is likely to be systematically limited.

Given the very clean and simple final state,
one can optimistically expect to keep the total
systematic error at 0.5\% level. This will include
the absolute luminosity determination (0.1\%),
tracking and Particle Identification efficiency (0.15\% per track for the 
tracking
and a similar number for the PID),
trigger efficiency, residual background.

With the above systematic error, the DELPHI precision
can be improved by a factor of 8 for the cross-section
measurement and the sensitivity to the anomalous
magnetic moment would be improved by a similar factor.
Adding other final states (e.g. $e-\rho$ and $\mu-\rho$)
is not expected to provide a large improvement
of the total error, since the measurement will be
systematically dominated. However, those channels
will have partially independent systematic uncertainties,
providing important cross-checks and certain reduction
of the overall systematic errors.

\section {Photon hadronic structure}
\label{sectPSF}

The measurement of the photon structure function (PSF) has a long history. 
Although the photon is considered as a point-like particle, 
due to the quantum effects it can fluctuate to a quark pair or a rho meson. 
These two processes are usually described respectively as point-like 
and hadron contribution to the photon structure function. 
The latter  could be estimated in the framework of vector meson 
dominance model  \cite{Brodsky:1971ud} while the former 
was calculated  within quark model in leading order QCD 
corrections \cite{Witten:1977ju} and later in series of papers 
at NLO \cite{Bardeen:1978hg},\cite{Duke:1980ij} and 
NNLO accuracy \cite{Ueda:2007yh}.

The most recent measurements of PSF have been performed at LEP 
about 20 years ago, and since then  publications are scarce, 
the most novel review of state-of the-art 
could be find at  \cite{Berger:2014rva},\cite{Schienbein:2002wj},\cite{Sasaki:2018hud}.

The photon hadronic structure can be tested via the measurement
of inclusive hadron production in  gamma-gamma
collisions $\rm{e^+e^-} \rightarrow \rm{e^+e^-} \gamma \gamma^*
\rightarrow \rm{e^+e^- + hadrons}$. The high-virtuality
photon $\gamma^*$ is radiated off an electron which scatters
at relatively large angle and can be detected in the experimental setup
(``tagged electron'').
The second electron is usually scattered at a very small angle
and thus remain undetected (``untagged electron'').
The photon $\gamma$ radiated by the untagged electron
can be considered as quasi-real.

The overall reaction can be described as a deep inelastic
scattering (DIS) $\rm{e}\gamma$ of the tagged electron 
on the real photon. In such a scattering the hadronic nature
of the target photon is effectively revealed.

Pioneering  experiment of PSF $F_2$ measurement was done at DESY 
by PLUTO collaboration in 1981  \cite{Berger:1981bh}, 
within the range $Q^2\in(1,15)$ GeV$^2$ at an average 
beam energy of $15.5$ GeV. 
Since then  many experiments made contributions so the $Q^2$ range 
was enlarged up to 780GeV$^2$ at  
PLUTO \cite{Berger:1984xt},\cite{Berger:1986hr}, ALEPH \cite{Barate:1999qy},\cite{Heister:2003an} AMY \cite{Kojima:1997wg},\cite{Sahu:1995gj}, DELPHI \cite{DaSilva:2000ng},\cite{Abreu:1995xta}, JADE \cite{Bartel:1984cg}, L3 \cite{Achard:2005fw},\cite{Acciarri:1998ig} , OPAL \cite{Abbiendi:2002te}, TASSO \cite{Althoff:1986fi}, TOPAZ \cite{Muramatsu:1994rq}, TCP$/2\gamma$ \cite{Aihara:1986xw}. 

The most recent experiments have been done at LEP 
by collaborations ALEPH, L3 and OPAL with similar characteristics: 
center of mass energy $\approx 200$ GeV, 
integrated luminosity $\approx$ 600 pb$^{-1}$ while the minimum angle 
for detection in the luminosity calorimeter was $0.024$, $0.03$ 
and $0.033$ rad correspondingly.
Comparing this data with similar CEPC parameters, 
namely  c.m.s. energy 240 GeV, integrated luminosity 5 ab$^{-1}$ 
and minimal detection angle of about 1.9 degree ($0.03$ rad) 
one can naively estimate  the number of events at CEPC to be 
about $10^4$ times bigger than that at LEP.

The hadronic photon structure function $F^\gamma_2(x,Q^2)$  
can be extracted from the differential cross-section
$d\sigma/dQ^2dxdy$, where $Q^2$ is the virtuality of $\gamma^*$
and $x,y$ are the Bjorken scaling variables.

For the case of single-tagged events, when one of the photon is almost real (photon invariant mass $P\approx 0$), the general process  
cross-section $e^+e^-\to e^+e^-\gamma\gamma\to e^+e^-X$ 
\begin{gather}
\frac{d\sigma(ee\to ee X)}{d x d z dQ^2 dP^2}=f_{\gamma/e}
\frac{d\sigma(e\gamma\to e X)}{d x dQ^2},
\end{gather}
are factorized into  almost real photon luminosity 
function $f_{\gamma/e}$
\begin{gather}
f_{\gamma/e}=\frac{\alpha}{2\pi}(\frac{1+(1-z^2)}{z}\frac{1}{P^2}-\frac{2m_e^2z}{P^4}), ~~~ z={E_\gamma}/{E_{beam}}
\end{gather}
and deep inelastic proton-electron scattering 
cross-section \cite{Brodsky:1971ud}
\begin{gather}
\frac{d\sigma(e\gamma\to e X)}{d x dQ^2}=
\frac{2\pi\alpha^2}{x Q^4}((1+(1-y)^2) F_2^\gamma-y^2F_L^\gamma).
\label{gamCRSect}
\end{gather}
Neglecting the virtuality of the quasi-real photon, one has:
$Q^2 = 2E_{beam}E_{tag}(1 - {\rm cos}\theta_{tag})$, $x = Q^2/(Q^2+W^2)$,
$y = 1 - (1 + {\rm cos}\theta_{tag})E_{tag}/(2E_{beam})$,
where $W$ is the $\gamma \gamma^*$ invariant mass which
is experimentally measured as the mass of the hadronic system, $\theta_{tag}$ and $E_{tag}$
are scattering angle and energy of the tagged lepton.

Usually the experimentally accessible kinematic region 
corresponds to small values of $y$ ~~ ($y^2\ll1$), 
so the contribution of the term proportional to the longitudinal 
structure function $F_L^\gamma$ is negligible, 
and one can determine PSF $F_2^\gamma$ directly 
from the cross-section (\ref{gamCRSect}).

The measurement of the energy $E_{tag}$ and the polar angle 
$\theta_{tag}$ of the scattered electron is straightforward.
The most difficult part of the analysis is the reconstruction
of the hadronic invariant mass $W$. In addition to the
finite detector resolution and efficiency, the $W$ reconstruction
is also affected by the acceptance issue, since the hadronic
system is typically boosted along the beam axis and part 
of it remains undetected or poorly reconstructed.
A sophisticated unfolding procedure is required
to convert the measured visible invariant mass into the true one.
An additional difficulty will be the background
from overlapping interaction. This background has
little impact on the hard processes, but must
be carefully taken into account in the studies
of gamma-gamma collisions.

The scattered electron can be reconstructed either in the luminosity
monitor (probing the small $Q^2$ values with high statistics),
or in the forward electromagnetic calorimeter. In the latter case
the domain of high $Q^2$ values can be accessed.
The available range of very high $Q^2$ is limited by the low
statistics due to the steeply falling spectrum   
of scattering angles. Given the unprecedented luminosity,
the future $e^+e^-$ colliders will be able to study the photon 
structure function
in the high $Q^2$ domain which was never accessed by
other experiments.

At LEP2 the explored range of $Q^2$ was limited to 
about $10^3$ GeV$^2$, mainly due to the available statistics.
At CEPC the collision energy will be comparable to LEP2.
However, the huge statistics expected at the future 
colliders (several orders of magnitude increase)
allows one to explore the kinematical regions that could not
be accessed by the past experiments due to the limited
statistics. Among them, the following can be mentioned:

\begin{itemize}
	
	\item Measurement of the photon structure function  
	at the very high virtualities $Q^2$, corresponding
	to the very large electron scattering angles.
	
	\item Double-tagged events, where both beam particles are detected.
	In this situation the whole event is fully reconstructed,
	which will dramatically reduce the systematic uncertainty
	at the expense of the low available statistics. 
	
	This case of deeply virtual target photons $Q^2\gg P^2\gg\Lambda_{QCD}$ is very interesting because it  is purely perturbative and allows one to compare the experimental values with absolute  QCD predictions.

\end{itemize}

\section {Conclusions}
In this paper we study several interesting topics of two-photon physics 
that could be considered at future electron-positron colliders, 
CEPC and FCC-ee. As a reference we take the planned CEPC set-up 
with c.m.s. energy $\sqrt{s}=240$ GeV, 
integrated luminosity 5.6 $ab^{-1}$ and assume
that a forward electromagnetic calorimeter
will cover the polar angles down to
$1.9$ degrees. Our qualitative results
can be applied to FCC-ee project, taking into account
its similar expected performance.  

We have considered a rather limited set of topics of two-photon physics, 
namely quarkonium  spectroscopy, Higgs and tau pair  production, 
photon hadronic structure. Our estimations are based
on a conservative approach and  without any doubt
the quantitative characteristics  can be surpassed
in the real experiments.
New  physics problems like searching for MSSM heavy Higgs,  
anomalous top quark interactions,
new physics state searching  will be considered elsewhere.

In Sect. \ref{qarkSect} we consider quarkonium two-photon physics 
in three possible detection modes: no-tag, single- and double tag.  
We show that using low-angle calorimeter (1.9 degree minimal 
detection angle) could drastically  improve statistics of 
tagged events by two orders of magnitude. 
In no-tag mode we expect about $10^7$-$10^8$ quarkonium  
and $10^5$ bottomonium events. Taking into account the efficiency 
of detection about $10\%$ and considering single-tag mode 
for better reconstruction of final state and reducing background events, 
we estimate about $10^2-10^4$ registered events for charmonium states. 
For bottomonium  we could expect about  $4\times10^2$ of  
low-state $\eta_b$  and possibly see some events or make upper bound   
for higher states.

We find (Sect. \ref{higgsSect}) that photoproduction 
of Higgs can be observed with
total  cross-section about $0.25$ fb and luminosity of collider 5.6 $ab^{-1}$.
Possible background issues from non-resonant $\gamma\gamma$ collisions 
and $ZZ\to H$ were discussed and it was shown that by choosing 
appropriate selection cuts 
one could achieve more than $4\sigma$ signal significance 
for the $H\to bb$ channel.

The tau pair production cross-section and anomalous magnetic moment 
measurement  can be improved up to a factor 8 compared with the 
precision of DELPHI experiment at LEP (Sect. \ref{tauSect}).

Statistical error of photon structure function measurement 
(Sect. \ref{sectPSF}) could be improved by about two orders 
of magnitude compared to the LEP experiments.
Also we expect that due to the high luminosity of the future
colliders it will be possible to perform measurements of PSF 
at very high virtualities $Q^2$. 

More detailed analysis and thorough simulation  of events, 
background, systematic and statistical errors of  physical 
quantities to measure  can be estimated after exact 
characteristics of planned colliders and detectors will be known.

\section{Acknowledgements}

Authors are grateful to prof. Li Haibo for fruitful
discussions which inspired this work.
The work of V.V.B. was supported in part by the Heisenberg-Landau Program.

\end{document}